\documentclass{elsart}
\usepackage[square,comma]{natbib}
\usepackage{epsfig}
\begin{document}
\runauthor{Kowalewski, Jackson}
\begin{frontmatter}
\title{Track Fit Hypothesis Testing and Kink Selection using
Sequential Correlations}
\author{Robert V. Kowalewski\thanksref{Principal}} and 
\author{Paul D. Jackson}
\address{Department of Physics and Astronomy,\\ 
University of Victoria, P.O. Box 3055,\\ 
Victoria, B.C. V8W 3P6, Canada} 
\begin{abstract}
Deviations between the form of trajectory assumed in a fit to a set of
measurements and the actual form of the trajectory can give rise to
sequential correlations in the residuals from the fit.  These
correlations can provide a more powerful goodness-of-fit test than
that based on the minimum ${\mathrm{\chi^2}}$ from a least squares
fit.  The use of this additional information is explored in the
context of several common trajectory errors (e.g. decays in flight)
encountered in charged particle tracking.
\end{abstract}
\begin{keyword}
charged particle tracking; hypothesis testing; statistical analysis;
kink finding; track fitting\\
PACS code: 07.05.Kf
\end{keyword}
\thanks[Principal]{Email address: kowalews@uvic.ca}

\end{frontmatter}

\section{Introduction}
Trajectory fitting aims to determine a set of fitted parameters and to
test the validity of the trajectory hypothesis.  Both of these
questions are usually addressed by minimizing a ${\mathrm{\chi^2}}$
constructed from the squares of the deviations between the
measurements and the parameterized trajectory.  The value of
${\mathrm{\chi^2}}$ at the minimum is used to test the adequacy of the
fitted hypothesis.\cite{Cowan} The ${\mathrm{\chi^2}}$ test, however,
explicitly ignores correlations amongst the residuals\footnote{The
term residual in this paper refers to the signed distance of closest
approach between a measurement and the fitted trajectory divided by
the uncertainty assigned to the measurement.} from the fit.  Such
correlations arise naturally in trajectory fitting, where adjacent
measurements are in causal order.  Deviations from the expected
trajectory due to either a discrete change at some point (e.g. a
scattering or decay in flight) or to a continuous parameter change
(e.g. dE/dx energy loss or magnetic field anomalies) introduce
correlated shifts in the positions of all subsequent measurements.
The ${\mathrm{\chi^2}}$ test is not very sensitive to these trajectory
deviations when they are small on the scale of the measurement errors,
since it considers only the squares of the residuals.

This paper introduces new quantities for hypothesis testing that are
applicable to any fits for which an ordering variable (e.g. time) can
be identified.\footnote{Not all fits are in this category: e.g. one
cannot in general define a useful ordering variable for vertex fits.}
The mean correlation of ordered sets of residuals (nearest neighbor,
next nearest neighbor, etc.) are used for this purpose.  These
correlations are essentially independent of ${\mathrm{\chi^2}}$, and
test the assumption that the measurements are mutually independent
(or, more generally, that the correlations amongst the measurements
are properly accounted for in the fit).  It should be noted
that the presumption of the independence of the measurements once
known sources of correlation (e.g. multiple Coulomb scattering) are
taken into account is also present in sequential fitting methods such
as the Kalman Filter\cite{Kalman,Fruhwirth}, and the correlation test
developed here can be applied to the output of such a fit.

The power of these correlation statistics for hypothesis testing is
studied as a function of several trajectory deviations that arise
naturally in charged particle tracking.  For simplicity, the
trajectories studied are circular arcs, corresponding to the
projection of charged particle trajectories onto a plane transverse to
a uniform axial magnetic field.  The following sections describe the
correlation variables, the simulation used to measure their
effectiveness, and the improvement in discrimination between true and
false hypotheses beyond what can be achieved using ${\mathrm{\chi^2}}$
alone.

\section{Description of correlation variables}
The degree to which nearby measurements are correlated can be gauged
by considering the mean correlation as a function of the distance
between measurements.  Using the standard correlation
estimator\cite{Cowan} to form an average correlation 
\begin{equation}
  {\mathrm{
   r_k = 
     { \sum_i
         \frac{\delta_i\, \delta_{i+k}}{\sigma_i\, \sigma_{i+k}}}
\ /\ { \sqrt{\left(\sum_i \frac{\delta_i^2}{\sigma_i^2}\right)
             \left(\sum_i \frac{\delta_{i+k}^2}{\sigma_{i+k}^2}\right)}}
   }}
\end{equation}
as a function of the distance between measurements gives fairly good
discrimination between true and false trajectory hypotheses.  The sums
run from 1 to ${\mathrm{N-k}}$, ${\mathrm{N}}$ is the number of
measurements on the trajectory, ${\mathrm{\delta_i}}$ is the signed
distance to the fitted trajectory for measurement ${\mathrm{i}}$,
${\mathrm{\sigma_i}}$ is the estimated uncertainty of measurement
${\mathrm{i}}$, and ${\mathrm{k}}$ is the correlation distance:
${\mathrm{k\in [1,N-1]}}$.  However, the following combination,
\begin{equation}
  {\mathrm{
   \rho_k = 
   { \sum_i w_i
     \frac{2\, \delta_i\, \delta_{i+k}}{\delta_i^2+\delta_{i+k}^2}}
 \ /
 \ { \sum_i w_i }
 \ \,;\ \ w_i = \frac{\delta_i^2+\delta_{i+k}^2}{\sigma_i^2+\sigma_{i+k}^2}
   }}
\end{equation}
\begin{figure}[h*tbp]
\begin{center}
\resizebox{130mm}{122mm}{\includegraphics*[0,0][530,530]{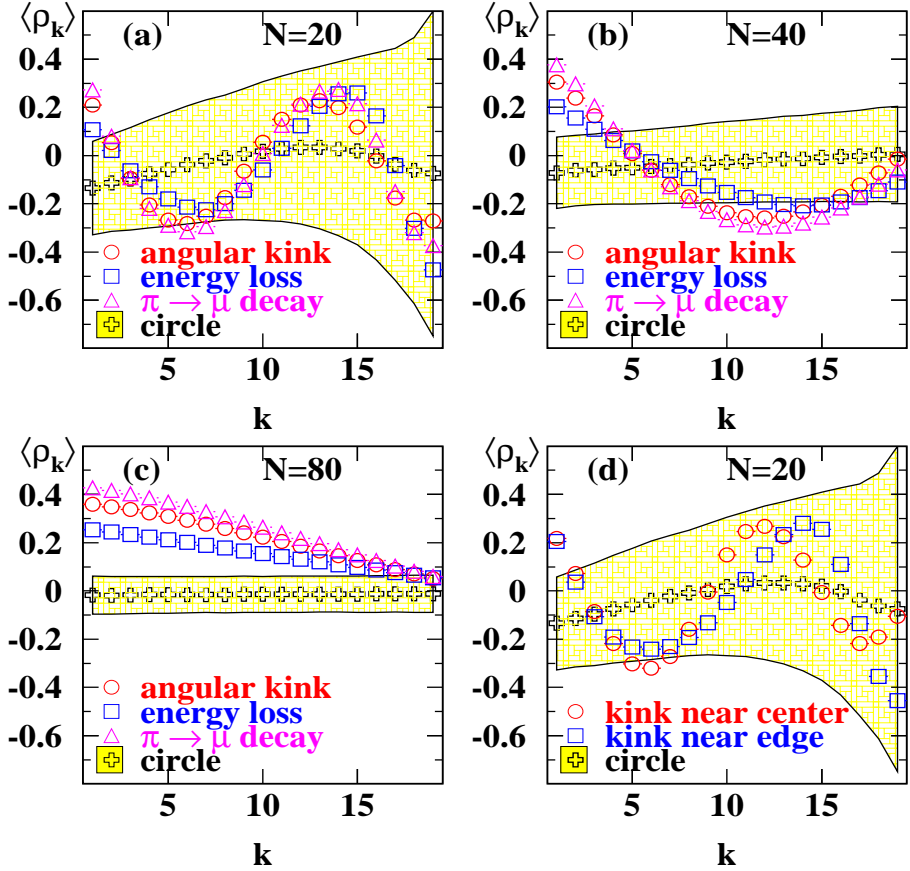}}
\caption{The expectation value of the correlation variable
${\mathrm{\rho_k}}$ is shown as a function of the correlation length
${\mathrm{k}}$.  The first 19 ${\mathrm{\langle\rho_k\rangle}}$ are
shown for 20 (a), 40 (b) and 80 (c) measurements.  The different
symbols correspond to specific generated trajectories; the fitted
trajectory is a circle in each case.  The shaded region indicates the
r.m.s. of the ${\mathrm{\rho_k}}$ distribution for the circle
trajectory.  (d) shows ${\mathrm{\langle\rho_k\rangle}}$ for angular
kinks occurring at different locations.
\label{fig:one}}
\end{center}
\end{figure}
motivated by considering the measure
${\mathrm{(\delta_i-\delta_{i+k})^2\,/\,(\sigma_i^2+\sigma_{i+k}^2)}}$,
gives slightly better discrimination.  This is because the correlation
sought is in the actual distances ${\mathrm{\delta}}$ from the fitted
trajectory, not in the residuals ${\mathrm{\delta/\sigma}}$.  The
weight factor emphasizes those pairs of measurements with significant
deviations.  These correlation measures satisfy ${\mathrm{\left|
\rho_k \right| \le 1}}$ for all ${\mathrm{k}}$.  For uncorrelated
measurements the expectation values of the ${\mathrm{\rho_k}}$ are
close to zero.  Negative correlations are introduced by the trajectory
fit; they are small provided the number of fitted parameters is much
smaller than the number of measurements.

Fig.~\ref{fig:one}(a-c) shows the expectation
value\footnote{Calculated numerically using the simulation described
below.} of each ${\mathrm{\rho_k}}$ as a function of the correlation
distance k (for ${\mathrm{k}}$ up to 19) for true circle trajectories
and for three common trajectory deviations: a discrete angular kink,
an uncorrected continuous energy loss and the decay in flight of a
pion to a muon.  The locations of the angular kink and decay in flight
were uniformly distributed along the trajectory.  The fitted
trajectory in each case was a circular arc.  The magnitudes of the
${\mathrm{\langle\rho_k\rangle}}$ for the incorrect hypotheses depend
on the particular choice of parameters in the simulation.  The number
and location of the crossing points between
${\mathrm{\langle\rho_k\rangle}}$ for the incorrect trajectory
hypotheses and ${\mathrm{\langle\rho_k\rangle}}$ for the correct
hypothesis are very similar for the trajectory deviations studied
here.  Fig.~\ref{fig:one}(d) compares a sample with a discrete angular
kink located near the center of the measurement region with a sample
where the kink is located closer to the edge of the measurement
region.  The kink location clearly has an impact on the behavior of
${\mathrm{\langle\rho_k\rangle}}$, most notably for large
${\mathrm{k/N}}$.

Using all of the ${\mathrm{\rho_k}}$ values gives the best
discrimination between the correct hypothesis and a particular type of
incorrect hypothesis, but does not provide optimal discrimination for
all types of incorrect hypothesis as can be seen by considering the
different shapes (note in particular the locations of the extrema) of
the trajectory hypotheses shown in Fig.~\ref{fig:one}.  Furthermore,
most of the discrimination power is concentrated at small
${\mathrm{k}}$, since the r.m.s. of the ${\mathrm{\rho_k}}$
distributions expected for the correct hypothesis are smallest
there.\footnote{The number of correlation measures summed for a given
${\mathrm{k}}$ is ${\mathrm{N-k}}$, giving a statistical error
proportional to ${\mathrm{1/\sqrt{N-k}}}$.}  These considerations,
along with the essentially linear behavior of the difference
${\mathrm{\langle\rho_k\rangle_{false} -
\langle\rho_k\rangle_{true}}}$ as a function of the correlation
distance ${\mathrm{k}}$ for small ${\mathrm{k/N}}$, lead us to
the following test statistic:
\begin{equation}
 {\mathrm{
  \lambda = \sum_{k=1}^{L} C_k \rho_k \ with\ 
    C_k = \frac{2}{L(L-1)}(L-k)\ \ and\ \sum_{k=1}^{L} C_k = 1
 }}.
\end{equation}
Choosing ${\mathrm{L}}$ as the nearest integer to ${\mathrm{N/8}}$
was found to give good sensitivity to the trajectory deviations studied.

\section{Description of simulation}
The sensitivity of ${\mathrm{\lambda}}$ to trajectory deviations was
studied using a simple simulation.  Charged particles were generated
and tracked through a uniform axial magnetic field and measured points
were generated in the plane orthogonal to the magnetic field
direction. The measurement uncertainty was taken as
Gaussian. Trajectories were generated with parameters typical of
charged particle tracking detectors\cite{babartdr}:
\begin{itemize}
\item average hit resolution: 150 ${\mathrm{\mu m}}$;
\item variation in resolution: factor of 5 between the best and worst 
measured points;
\item radial difference between first and last measurement layer: 54 cm;
\item number of measurements: varied from 20 to 160;
\item magnetic field: axial field of magnitude 1.5 Tesla;
\item initial particle momentum between 0.5 GeV/c and 5 GeV/c.
\end{itemize}
The measurements were uniformly distributed along the trajectories,
the hit efficiency was unity and no noise hits were generated.  The
location of discrete trajectory deviations (angular kink or particle
decay) was randomly distributed along the trajectory.  The generated
points were fitted to a circle assuming perfect pattern recognition,
i.e. all generated measurements were used in the fit.

\begin{figure}[h*tbp]
\begin{center}
\resizebox{85mm}{80mm}{\includegraphics*[0,0][530,530]{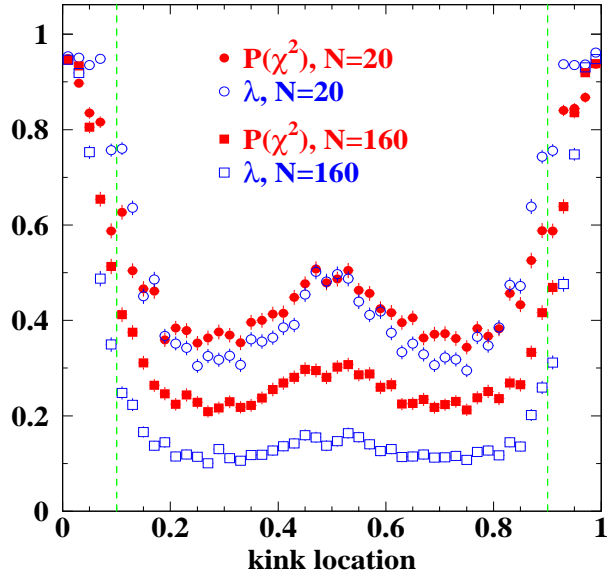}}
\caption{The survival rate as a function of the location of a
discrete angular kink within the measurement volume.  The cuts on
${\mathrm{\lambda}}$ and ${\mathrm{P(\chi^2)}}$ are set to give 95\%
efficiency for true circular trajectories.  The dashed lines
bound the region used for subsequent study of the power of each
variable.  \label{fig:two}}
\end{center}
\end{figure}
\section{Results}
The generated data were used to study the discrimination power of
${\mathrm{\lambda}}$ and of ${\mathrm{\chi^2}}$ as a function of
specific trajectory deviations.  Fig.~\ref{fig:two} shows the fraction
of trajectories surviving a cut
on ${\mathrm{\lambda}}$ or on the ${\mathrm{\chi^2}}$ probability
${\mathrm{P(\chi^2)}}$ as a function of the position of a discrete
angular kink, where the kink angle was uniformly distributed between
${\mathrm{\pm 0.02}}$ radians.  The cuts were set to give 95\%
efficiency for true circular trajectories.  There is little
discrimination power for kinks occurring at either end of the
measurement region.  Based on this, only those trajectory deviations
occurring within a fiducial region consisting of the central 80\% of
the measurements were selected for subsequent study.

\begin{figure}[h*tbp]
\begin{center}
\resizebox{130mm}{122mm}{\includegraphics*[0,0][530,530]{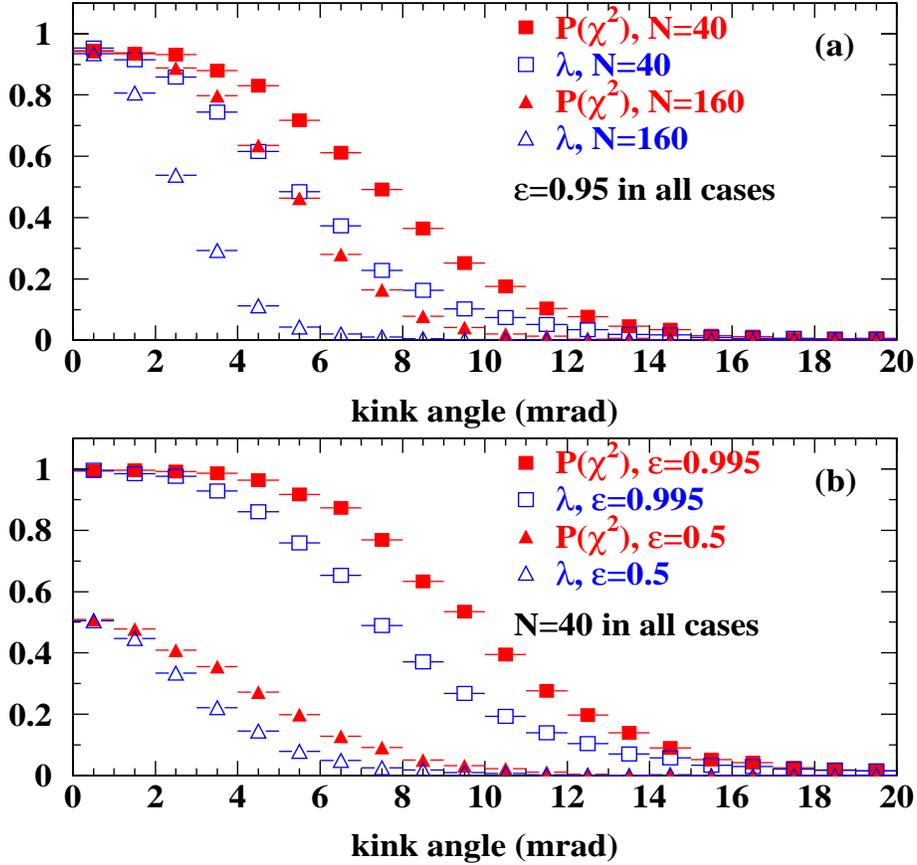}}
\caption{The survival rate as a function of the size of a discrete
angular kink within the fiducial region.  The behavior for different N
is shown in (a) for a cut that gives an efficiency ${\mathrm{\epsilon}}$
of 95\% for true circular trajectories.  The curves in (b) show the
effect of varying the cut on ${\mathrm{\epsilon}}$.
\label{fig:three}}
\end{center}
\end{figure}

\begin{figure}[h*tbp]
\begin{center}
\resizebox{130mm}{122mm}{\includegraphics*[0,0][530,530]{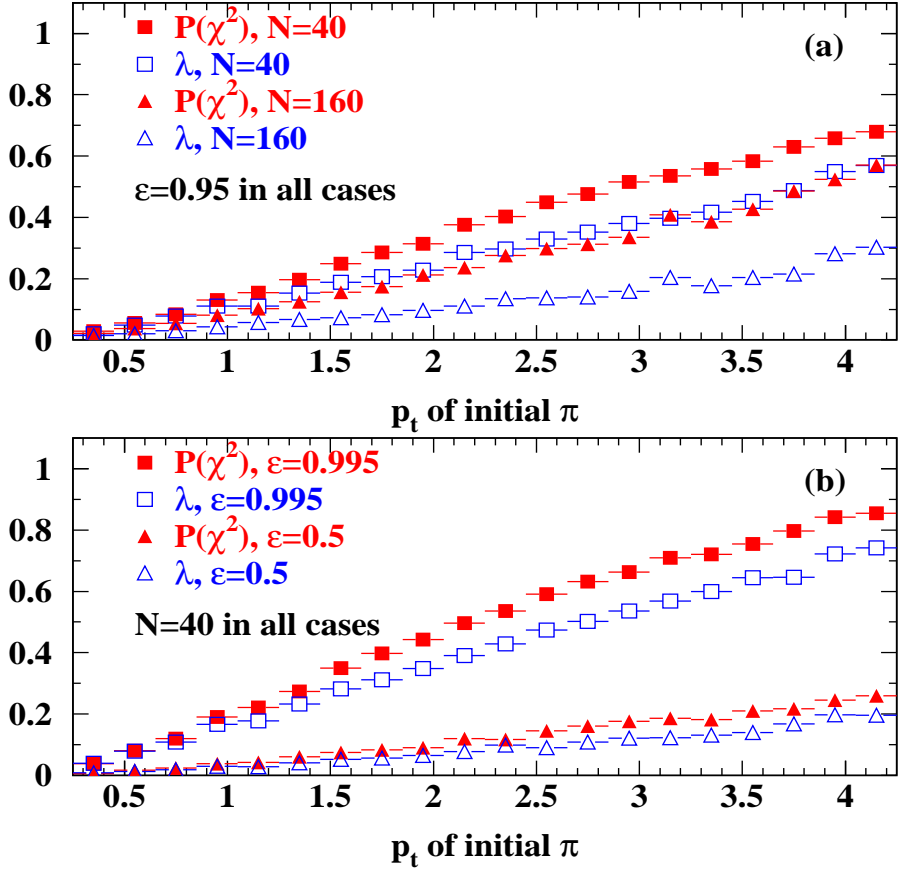}}
\caption{The survival rate for ${\mathrm{\pi\rightarrow \mu}}$ decays
occurring in the fiducial region as a function of the initial momentum
of the pion in the plane transverse to the magnetic field.  The
behavior for different N is shown in (a) for a cut that gives an
efficiency ${\mathrm{\epsilon}}$ of 95\% for true circular trajectories.
The curves in (b) show the effect of varying the cut on
${\mathrm{\epsilon}}$.
\label{fig:four}}
\end{center}
\end{figure}

The fraction of these selected trajectories surviving a cut that gives
95\% efficiency for true circular trajectories is shown as a function
of the size of the angular deviation in Fig.~\ref{fig:three}.  The
correlation variable ${\mathrm{\lambda}}$ is more powerful than
${\mathrm{P(\chi^2)}}$ provided ${\mathrm{N>20}}$ and becomes
relatively more powerful as the number of measurements increases.
This is to be expected, since the physical correlation length sampled
by ${\mathrm{\lambda}}$ is of the order of ${\mathrm{\frac{1}{8}}}$ of
the track length, and increasing the density of measurements allows a
more precise determination of the correlation.  Similar behavior is
observed for different choices for the efficiency for true circular
trajectories.  The power of ${\mathrm{\lambda}}$ and
${\mathrm{P(\chi^2)}}$ to discriminate ${\mathrm{\pi\rightarrow \mu}}$
decays from true circular trajectories is shown in
Fig.~\ref{fig:four}.  Note that the decay angle between the
${\mathrm{\pi}}$ and ${\mathrm{\mu}}$ does not in general lie in the
measurement plane, and that the momentum of the muon will be smaller
than that of the pion.  The correlation variable ${\mathrm{\lambda}}$
is again a significantly better discriminant than is
${\mathrm{P(\chi^2)}}$.

\begin{figure}[h*tbp]
\begin{center}
\resizebox{130mm}{122mm}{\includegraphics*[0,0][530,530]{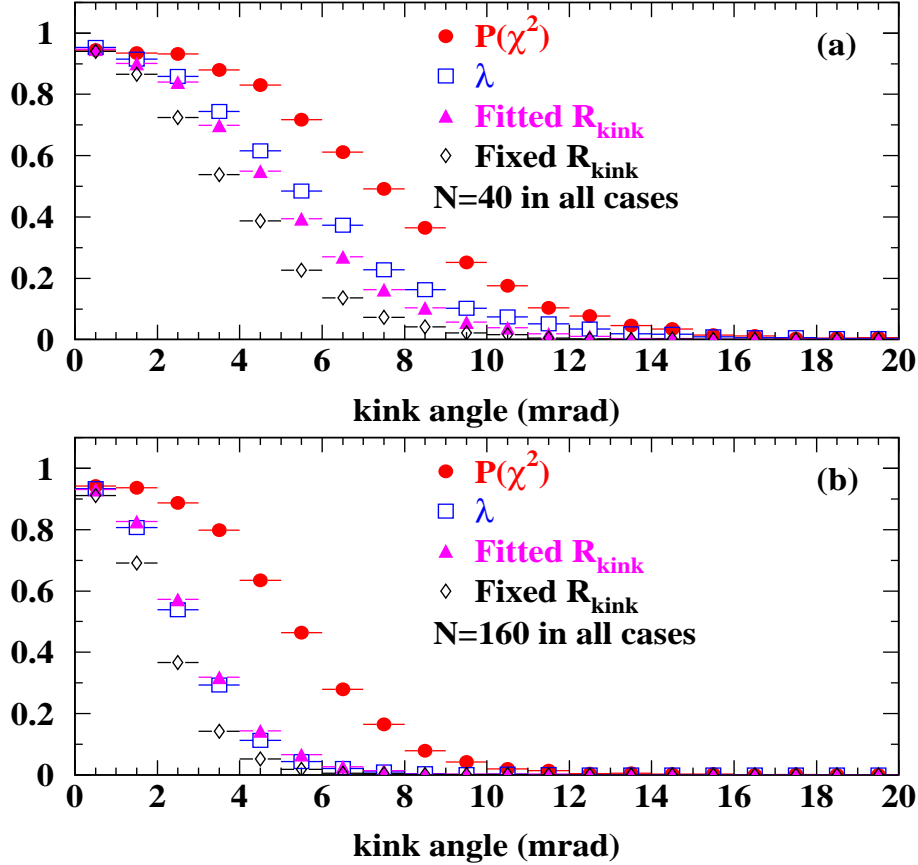}}
\caption{The survival rate as a function of the size of a discrete
angular kink for 40 (a) and 160 (b) measurements.  The cuts on
${\mathrm{\lambda}}$ and ${\mathrm{P(\chi^2)}}$ are set to give
95\% efficiency for true circular trajectories.  Also shown are results
from an optimal technique (see text) with the kink location either
fixed to the generated position (``fixed ${\mathrm{R_{kink}}}$'') or
determined by the fit (``fitted ${\mathrm{R_{kink}}}$'').
\label{fig:five}}
\end{center}
\end{figure}

The extent to which the discrimination afforded by
${\mathrm{\lambda}}$ is optimal was studied.  The measurements along
trajectories generated with a discrete angular kink were fitted using
both the correct hypothesis, i.e. two circular arcs of constant
curvature with an angular deviation at a point, and using the nominal
(incorrect) hypothesis of a single circular arc.  The ratio of the
resulting ${\mathrm{\chi^2}}$ probabilities
${\mathrm{P_{circle}(\chi^2)/P_{correct}(\chi^2) }}$ is an optimal
discriminant variable in this case.\footnote{In practice, trajectory
deviations of several different types may be present in a sample, so
fitting for one particular type of deviation will not be optimal.}
The position of the angular kink was taken as either the true position
of the generated deviation (denoted ``fixed ${\mathrm{R_{kink}}}$'' in
Fig.~\ref{fig:five}) or as the best fit value after considering all
potential kink positions (denoted ``fitted ${\mathrm{R_{kink}}}$'' in
Fig.~\ref{fig:five}).  The kink position will in general be unknown,
so the ``fitted ${\mathrm{R_{kink}}}$'' curve is in practice the best
one can achieve.  As is seen in Fig.~\ref{fig:five}, the correlation
variable ${\mathrm{\lambda}}$ gives discrimination that approaches the
optimal value as the number of measurements increases.

\section{Discussion}

The correlation variables ${\mathrm{\rho_k}}$ introduced here test the
assumption that a set of measurements are mutually independent once
known sources of correlation have been taken into account.  Use of the
simple combination of ${\mathrm{\rho_k}}$ introduced here as
${\mathrm{\lambda}}$ leads to an improvement in the sensitivity for
detecting small trajectory deviations relative to that achievable
using ${\mathrm{P(\chi^2)}}$.  This additional goodness-of-fit test is
independent of the ${\mathrm{\chi^2}}$ test, and can be applied to the
residuals from both traditional least squares fits and from Kalman
filter fits.  The improved sensitivity to small deviations can be
qualitatively understood by recognizing that ${\mathrm{\chi^2}}$ has a
quadratic dependence on individual deviations, while
${\mathrm{\lambda}}$ has a linear dependence.  For larger deviations
(not shown here) ${\mathrm{P(\chi^2)}}$ becomes a more powerful
discriminant than ${\mathrm{\lambda}}$, as expected.  For the
deviations studied, the gain from combining the ${\mathrm{\lambda}}$
and ${\mathrm{P(\chi^2)}}$ tests was negligible.

The ${\mathrm{\lambda}}$ and ${\mathrm{\chi^2}}$ tests have different
dependencies on the input to the fit.  The ${\mathrm{\chi^2}}$ test is
sensitive to the scale of the assigned measurement errors
${\mathrm{\sigma_i}}$ and can be compromised by mis-estimates of and
non-Gaussian contributions to the measurement errors.  The
${\mathrm{\lambda}}$ test is insensitive to the scale of the
${\mathrm{\sigma_i}}$.  It is, however, sensitive to correlations
introduced in calibration procedures.  The extent to which this is a
practical problem in using the ${\mathrm{\lambda}}$ test is a function
of detector design and calibration.  In particular we expect the
${\mathrm{\lambda}}$ test to be most useful in devices where the
effect of calibrations is randomized over the measurements on a
trajectory (e.g. in small cell drift chambers, where the drift
direction changes layer by layer), and to be less effective in devices
where coherent effects dominate (e.g. in detectors employing a jet
cell design).

Potential uses for the ${\mathrm{\lambda}}$ test in charged particle
tracking involve selecting decays in flight and enabling high quality
track samples (with reduced non-Gaussian tails on the track parameter
resolutions due to trajectory deviations) to be selected.  Given the
ease with which it can be calculated, a test on ${\mathrm{\lambda}}$
might serve as a filter for selecting tracks on which more
computationally intensive tests\cite{Fruhwirth,Cousins} will be
performed.  The correlations measured by the ${\mathrm{\rho_k}}$
parameters may find use in a broader range of applications.


\vspace{7mm}

\section{Acknowledgments}

The authors would like to thank Dr. Michael Roney for useful
discussions and to acknowledge Louis Desroches, whose work with
one of us (Kowalewski) on run test variables was a precursor to the
present work.


\end{document}